\documentclass[preprint,12pt]{elsarticle}

\usepackage{graphicx}
\usepackage{soul,color}
\usepackage{amssymb}

\usepackage{lineno}

\usepackage{booktabs}

\journal{arXiv.org}

\begin{document}

\begin{frontmatter}


\title{The resumption of sports competitions after COVID-19 lockdown: The case of the Spanish football league.}



\author[madrid1,madrid2,madrid3,madrid4]{Javier M. Buld\'u}
\author[madrid1,madrid2,madrid3]{Daniel R. Antequera}
\author[madrid2,astro]{Jacobo Aguirre}

\address[madrid1]{Complex Systems Group, Universidad Rey Juan Carlos, Madrid, Spain}
\address[madrid2]{Grupo Interdisciplinar de Sistemas Complejos (GISC), Spain}
\address[madrid3]{Laboratory of Biological Networks, Center for Biomedical Technology, Universidad Polit\'ecnica de Madrid, Madrid, Spain}
\address[madrid4]{Institute of Unmanned System and Center for OPTical IMagery Analysis and Learning (OPTIMAL), Northwestern Polytechnical University, Xi'an 710072, China}
\address[astro]{ Centro de Astrobiolog\'{\i}a (CSIC-INTA), Ctra. de Ajalvir km 4, 28850, Torrej\'on de Ardoz, Madrid, Spain.}

\begin{abstract}

In this work, we present a stochastic discrete-time $SEIR$ {\em Susceptible-Exposed-Infectious-Recovered} 
model adapted to describe the propagation of COVID-19 during a football tournament. Specifically, 
we are concerned about the re-start of the Spanish national football league, {\it La Liga}, 
which is currently --May 2020-- stopped with $11$ fixtures remaining. 
Our model includes two additional states of an individual, confined and quarantined, which are reached when an individual 
presents COVID-19 symptoms or has undergone a virus test with a positive result. 
The model also accounts for the interaction dynamics of players, considering three different
sources of infection: the player social circle, the contact with his/her team colleagues during training sessions, and 
the interaction with rivals
during a match. Our results 
highlight the influence of the days between matches, the frequency of virus tests and their sensitivity
on the number of players infected 
at the end of the season. 
Following our findings, we finally present a variety of strategies to minimize
the probability that COVID-19 propagates in case the season of {\it La Liga} was re-started after the current lockdown.

\end{abstract}

\begin{keyword}
Epidemics \sep SEIR \sep COVID-19 \sep Sports  

\end{keyword}

\end{frontmatter}


\section{Introduction}
\label{S:Intro}

The propagation of the virus SARS-CoV-2 officially started at the beginning of December 2019 in Wuhan (China), where the first 
COVID-19 victim was 
diagnosed with a new type of coronavirus. The virus first spread over different states in China before reaching other countries.
On March 11th 2020, the World Health Organization (WHO) declared COVID-19 a pandemic, pointing to more than $118000$ cases of 
the coronavirus illness in over 110 countries around the world \cite{who_pandemic}.
The evolution of the pandemic, which is (in May 2020) still affecting many countries worldwide, has been a matter of debate, since different 
strategies can be adopted
to mitigate the spreading of COVID-19, some of them with unclear or unpredictable consequences. Due to the novelty of this 
unforeseen pandemic, the use of mathematical models is being
extremely useful to predict the dynamics of the coronavirus spreading and the effects of different policies on the eventual 
reduction of the number of affected individuals.

Despite there are different approaches for modelling the pandemics, both continuous-time and discrete-time SIR-based models are probably the most extended approaches. 
The Susceptible-Infected-Recovered (SIR)
model was first proposed by Kermack and McKendrick in 1927 \cite{kermack1927}, and consists of a compartmental model where individuals are split into three different states: (i) Susceptible (S),
when they are sane, (ii) Infected (I), when they have the virus and (iii) Recovered (R). More sophisticated models include more possible states, such as Deceased (D)
in the SIRD model \cite{matadi2014} or Exposed (E) in the SEIR model \cite{li1995}. The latter model has been extensively applied to describe the exponential growth of the number
of individuals infected by SARS-CoV-2, the effects of quarantine and confinement measures and, ultimately, to evaluate an adequate way of leaving confinement measures without
increasing the risk of a second outbreak \cite{peng2020,radulescu2020,hou2020,rovetta2020,debrouwer2020,ngonghala2020,lopez2020}.
For example, Peng {\it et al.} collected the epidemic data from five different Chinese regions and estimated the effects of the quarantine over all of them, forecasting the decrease
of the number of infected individuals region by region \cite{peng2020}. On the other hand, Radulescu {\it et al.} introduced a compartmental model consisting of dividing the population
into age groups and analyzing how the number of infected individuals was related to each age group. With this model, the effects of several social measures were simulated
 (closing campuses, schools or restaurants), showing different impacts at each age group \cite{radulescu2020}. 
Regarding the vast (and recent) scientific literature about the SEIR
 model applied to COVID-19, we can remark one of its significant merits: It can be easily adapted to describe a diversity of scenarios.

In this manuscript, we present a discrete-time SEIR-type mathematical model that describes the spreading of the coronavirus during a sports competition. 
The motivation behind our study is that there has been a lively debate about whether sports competitions that were not completed before the coronavirus crisis
should be re-started or, ultimately, cancelled \cite{halabchi2020,duarte2020,corsini2020}. 
On the one hand, it is not the first time that epidemic diseases have threatened sports competition. For example, as pointed in \cite{duarte2020},
the 2014 FIFA World Cup in Brazil overlapped with a period in which Dengue risk was close to its maximum at three cities where matches were carried out \cite{hay2013}. Furthermore,
attendants and players had to take special precautions due to Zika, a mosquito-transmitted disease. Despite the risks, the competition continued without significant problems regarding
the number of individuals infected by Dengue or Zika. On the other hand, many voices have claimed that sports competitions should be canceled, not only for the high risk of athletes
being infected during a competition but also due to the inability to be adequately treated in case of injury due to the saturation of hospitals \cite{corsini2020}.

However, to the best of our knowledge, this debate has not 
been confronted with mathematical models that describe
the propagation of SARS-CoV-2 between athletes.
Here, we are concerned about the eventual re-start of the Spanish national league, which is currently suspended with $11$ pending fixtures, and focus on 
the optimum strategies to minimize the propagation
of COVID-19 among the players in case the competition was re-started after the current lockdown. We designed a mathematical 
model that incorporates the interaction
of players during training sessions, leading to intra-club spreading, and during matches, responsible for inter-club contagions. 
Furthermore, we incorporated
the use of tests to evaluate its consequences in identifying and confining those players that already have been infected. The model, whose main parameters were
based on the scientific literature concerning the infection and recovery periods of COVID-19, could be easily adapted to describe other kinds of sports competitions just
by modifying the number of players and teams participating in the tournament.

\section{Methodology}
\label{S:Methods}

In SEIR models \cite{Newman:2010}, a disease propagates 
through a network of individuals whose dynamical state can be either Susceptible  (S, healthy and susceptible to be infected),
Exposed (E, infected but in the latent period --period from infection to infectiousness-- and therefore unable to infect other individuals), 
Infectious (I, infected and able to infect 
other individuals), and Removed (R, which includes (i) recovered individuals after having suffered the infection and therefore immune and (ii) deceased people).

\begin{figure}[!h]
\centering
 \includegraphics[width=\textwidth]{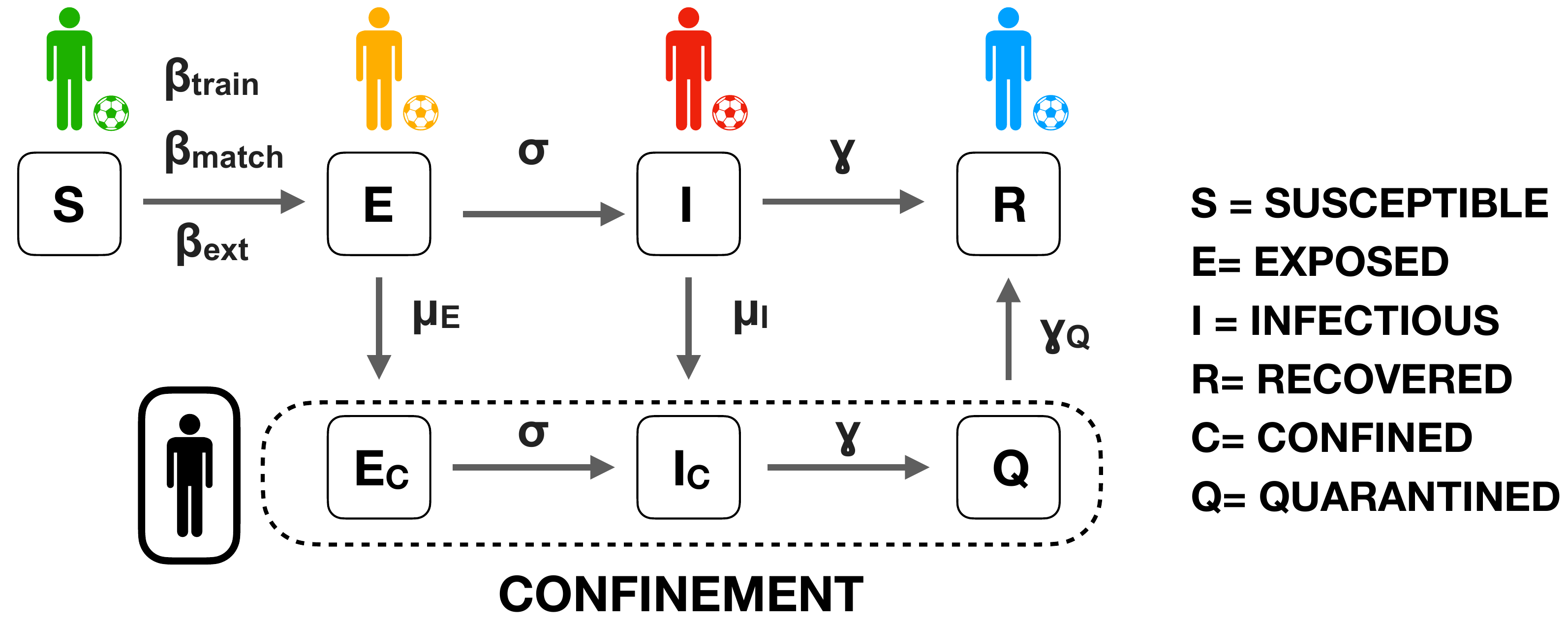}
\caption{ 
Schematic representation of the SEIR model adapted to a football competition. Players can be in different states:
Susceptible (S), Exposed (E), Infectious (I) and Quarantined (Q). In case players are detected to be infected by the virus, they
remain confined (indicated by the C suffix). After confinement, players undergo a quarantine
before being eligible to play again. Parameters $\beta_{train}$, $\beta_{match}$ and $\beta_{ext}$ account for the probability of 
becoming Exposed (E) during training, matches or externally (player social circle) respectively. 
Probability $\sigma$ describes the transition from  
Exposed to Infectious (I). Probability $\gamma$ controls the transition from Infectious to Recovered (R) or Quarantined (Q). Finally,
$\gamma_Q$ is related to the quarantine period a player must follow after recovery. 
\label{figura_modelo}
} 
\end{figure}

Figure \ref{figura_modelo} represents a sketch of our discrete-time model. The time is discretised in days, and every random event is
calculated once a day. The individuals (players from now on) 
can be infected at any time (that is, any day of the season)
from people different to the players (technical staff of the team, family, etc.) with a probability $\beta_{ext}$. 
The second source of infection occurs 
during the training period, where they can be infected from other players of their
own team with probability $\beta_{train}$.  Finally, during the matches, players are exposed to infection 
from the players of their own team and the adversary 
team with probability $\beta_{match}$. \footnote{Note that the infection probabilities 
$\beta_{ext}$, $\beta_{train}$ and  $\beta_{match}$ are different to the
infection rate $\beta$ typically used in SIR models, which is the average number of contacts per person per time 
multiplied by the probability of disease transmission in a contact between a susceptible and an infectious subject. While $\beta$ is a rate
(and can therefore be larger than 1), our parameters are probabilities (and $\leq 1$ in consequence).} 

Once a player has been infected and becomes exposed, he/she has a probability $\sigma$ of finishing 
the latent period and become infectious. Exposed and infectious players have, respectively, probabilities $\mu_{E}$ and $\mu_{I}$ of 
being detected as infected by COVID-19 via a virus test or because they show disease symptoms. If this is the case,
players will be confined at their homes remaining in two possible states: exposed E$_C$ or infectious I$_C$. 
Asymptomatic infectious players (belonging to class I, but not detected by virus tests), 
and confined infectious players, overcome the disease with probability 
$\gamma$. Note that confined players that have been recovered will remain quarantined (class Q) 
during a convalescence until they are prepared for playing again and become recovered (R) with probability $\gamma_{Q}$.

The days between virus tests $N_{test}$ and the days between matches $N_{match}$ are two critical variables 
for controlling the number of infected players during the championship, 
and therefore their influence in the model should be studied carefully. 
Note that the virus tests should be done in this context via Polymerase Chain Reaction (PCR) controls. 
The reason is that fast antibody or antigen detectors are only reliable more than a week after the infection, and in many cases even after the patient has already shown symptoms. 
This fact would allow the infectious (but not identified as infected) players to spread the virus for several days, 
making the control of the disease a hard task.

\subsection*{Modelling the Spanish national league}

Our model can be applied to a diversity of competitions related to team contact sports, 
but we have focused on the re-start of the Spanish
male national league. Therefore, we considered a competition with $M=20$ teams composed of $L=25$ players, 
the latter being the upper limit of players
that can be registered by a team in the competition. 
The generalization to {\it Liga Iberdrola} (Spanish feminine first division football league, with 16 teams), 
to the masculine or feminine football leagues of other countries, or even to
another team sports (such as basketball, handball, rugby, etc.) is direct. 
Every team plays a match every $N_{match}$ days (in particular in days that are multiples of $N_{match}$), 
and during the $N_{match}-1$ days in between the players train at their own stadiums. 
We supposed no resting days, as there
is a clear interest for finishing the leagues as soon as possible, but including them in the model is trivial. 
We represented the training dynamics of the players, and contacts between them, using social networks instead of mean-field contacts.
In this way, players' social networks during the team training followed a random structure of connections (different for each
team but maintained during all simulations)
and were generated using an Erd\"os-R\'enyi model \cite{Erdos:1959} 
with a probability $p=0.2$ of connecting two players. 
This was done to describe the internal 
professional and friendship dynamics that every player has during training times and also during lunch time, etc.
During training time, the infectious players (class I) might infect their neighbours in the social
network with probability $\beta_{train}$. During the match day, every infected player on the pitch can infect any 
other player of its own team or the adversary
with probability $\beta_{match}$ (here we used a mean-field approach due to the inevitable contact dynamics that players follow
during a match). Note that players cannot avoid voluntarily the contact with other players in the contest (with the exception,
perhaps, of celebrating a goal, that could be forbidden if necessary), 
and therefore the contagion probability during a match might be more significant than expected at first glance. Also, as a third infection
source, players
can be infected any day from their social circle with probability $\beta_{ext}$. 
Finally, in order to minimise the spreading of the disease, 
a virus test is done to all 500 players every $N_{test}$ days (in particular in days that are multiples of $N_{test}$, and before the 
match if it coincides with a match day). Players that yield a positive result are immediately confined.

\subsection*{Estimation of the model parameters}
\begin{table}[!h]

\begin{center}

\begin{tabular}{|c|c|}
 \hline
 Parameter  & Value \\
\hline 
\hline
$\beta_{train}$ (days$^{-1}$) & 1/10 \\
\hline
$\beta_{match}$ (days$^{-1}$)& 1/100  \\
\hline
$\beta_{exp}$ (days$^{-1}$) & 1/100000 \\
\hline
$\sigma$ (days$^{-1}$)& 1/3;  Refs. \cite{Wu:2020,Lin:2020}\\
\hline
$\gamma$ (days$^{-1}$)& 1/5; Refs. \cite{Wu:2020,Lin:2020}\\
\hline
$\gamma_Q$ (days$^{-1}$)& 1/5\\
\hline
$\mu_E$ (days$^{-1}$)& $\mu_I/3$\\
\hline
$\mu_I$ (days$^{-1}$)& [0,1], 0.9\\
\hline
$N_{test}$ (days)& [1-7]\\
\hline
$N_{match}$ (days)& [3-7]\\
\hline

\end{tabular}
\end{center}
\begin{flushleft}
\end{flushleft}
\caption{
Summary of the main parameters used in the model: Probability of being infected during the training period
$\beta_{train}$, during a match $\beta_{match}$ and from the player's social circle $\beta_{exp}$; 
latent period $\sigma^{-1}$, infectious period $\gamma^{-1}$ and quarantine period $\gamma_Q^{-1}$;
probability of being detected as exposed (by virus test) $\mu_{E}$ and as infectious (by virus test or by symptoms) $\mu_{I}$;
number of days between virus tests $N_{test}$ and matches $N_{match}$.
\label{tab01}}
\end{table}

There is a wide range of values in the recent literature regarding each of the parameters that define the 
different steps of the disease (see Table 1 for a summary of the parameters of the model).
The latent period $\sigma^{-1}$ is the average time from infection to infectiousness, the incubation period is the average time 
from infection to the appearance of the first symptoms, and the infectious period $\gamma$ is the average time that the patient is infectious. Depending on the virus, 
the latent period can be shorter or larger than the incubation period. In the case of COVID-19 the latent period is $1$ or $2$ 
days shorter on average than the incubation period, which makes it especially easy for the disease to spread among the population
during the time in which people are infectious but asymptomatic. 
Regarding the mean incubation period, in \cite{Lauer:2020} it was shown to be around 5 days, similar to that of SARS, and 
in \cite{Guan:2020} it was affirmed that it could be as short as four days. 
Note, however, that this quantity was not used in our model.

In \cite{Wu:2020}, it was used a mean latent period $\sigma^{-1}$ of 3 days and a mean infection period of $\gamma^{-1}=5$ days, 
based on the Wuhan data. We selected these values because they were also used in other more recent studies \cite{Lin:2020}. 
Note, however, that these are mean values: in \cite{Wolfel:2020} it was shown that the probability that
patients with mild symptoms infected other people was very low after a week from 
the appearance of symptoms, but these means that in mild cases of COVID-19 patients can be infectious for as much as $10$ days. 
Furthermore, we have fixed the quarantine period $\gamma_Q^{-1}$ to be five days, 
but varying slightly this quantity would not affect the results substantially.

The probability $\beta_{ext}$ of being infected during a day from the player's social circle
will slowly decrease as more and more individuals in the country recover from the disease, but for simplicity 
we have supposed it constant during the whole league, and one order of magnitude lower than the expected 
value $\beta_{ext}'$ based on available Spanish statistics. The reason is that
players will be for sure either quarantined during the rest of the league (and in that case $\beta_{exp}= 0$)
or at least their social life will be very restricted during that time. In order to obtain a plausible value for $\beta_{ext}$,
we have used the data resulted from a mass virus testing campaign developed during 
the first days of May 2020 to 62400 Spanish citizens belonging to 30000 different homes. The main result of that study
is that around 5\% of
the Spanish population ($2.3\times 10^6$ out of $46.9\times10^{6}$ citizens) has been infected during the pandemic, 
and this represents 10 times the detected cases so far (229540 on 14th 
May 2020).  
As at this date around 500 new infections were detected per day, we can extrapolate that $N_{inf}\approx500\times9$ 
were not detected infected cases, and therefore $N_{inf}/\gamma$ people would be infective during the latent period. In summary, 
$$\beta_{exp}=\beta_{exp}'/10= \frac{(N_{inf}/\gamma) R_0 \gamma}{10 N_{Spain}}\approx 10^{-5}\,,$$ where the latent period 
is $\gamma^{-1}=5$ days
and the basic reproductive number (i.e. the expected number of infections generated by one case in a population 
where all individuals are susceptible to infection) is, according to the Spanish health authorities, $R_0\approx 1$ at this time. 

On the other hand, there is not available experimental data 
to obtain precise values for the infecting probabilities $\beta_{train}$ and $\beta_{match}$, so we have 
fixed them at moderate values and checked that slight variations did not qualitatively change the results. 
In particular, as on average each player is in contact, during the training time, with a fraction $p=0.2$ 
of the total number of players in the team ($L=25$), a first-order basic calculation yields that 
he/she will infect around $p L \beta_{train}=0.5$ 
other players per training day, as far as all other players are susceptible. 
During a match, nonetheless, an infectious player can infect any of the other 21 players in the field, 
and will infect on average $21 \beta_{match}=0.21$ players per match (supposing again that the rest are susceptible).
In summary, 
at the beginning of {\it La Liga}, and in the improbable situation that an 
infected player skipped all virus tests,
he/she would infect around two other players during the latent period, 
and this quantity would decrease with time as more and more
players become infected and then recovered. We believe this is a plausible result taking into account that
$R_0 > 4$ at the beginning of the pandemic and $R_0\approx 1$ after two months of absolute quarantine of the whole population 
of the country, and players would have an intermediate situation with a 
controlled but not quarantined behaviour. 

The value of the probability $\mu_{I}$ of being detected as infectious, either because 
a player shows disease symptoms or because the virus test yields a positive result, 
has been considered to be within the window [0,1], being $0$ in case of not doing any test and being asymptomatic,
and $1$ when tests have $100\%$ sensitivity.
However, when the sensitivity of the test is not analysed, we considered a value of $0.9$ which is close to the 
typical one attributed to PCR tests.
Concerning 
the probability of detecting an exposed individual, we set it as $\mu_{E}=\mu_{I}/3$, i.e., three times less than detecting an infected individual through
the same test. 
The reason is that the viral load of an exposed individual is lower than that of an infectious one,
therefore reducing the probability of a positive test result.

\section{Results}
\label{S:Results}

\begin{figure}[!t]
\centering
\includegraphics[scale=0.35,angle=0,clip=]{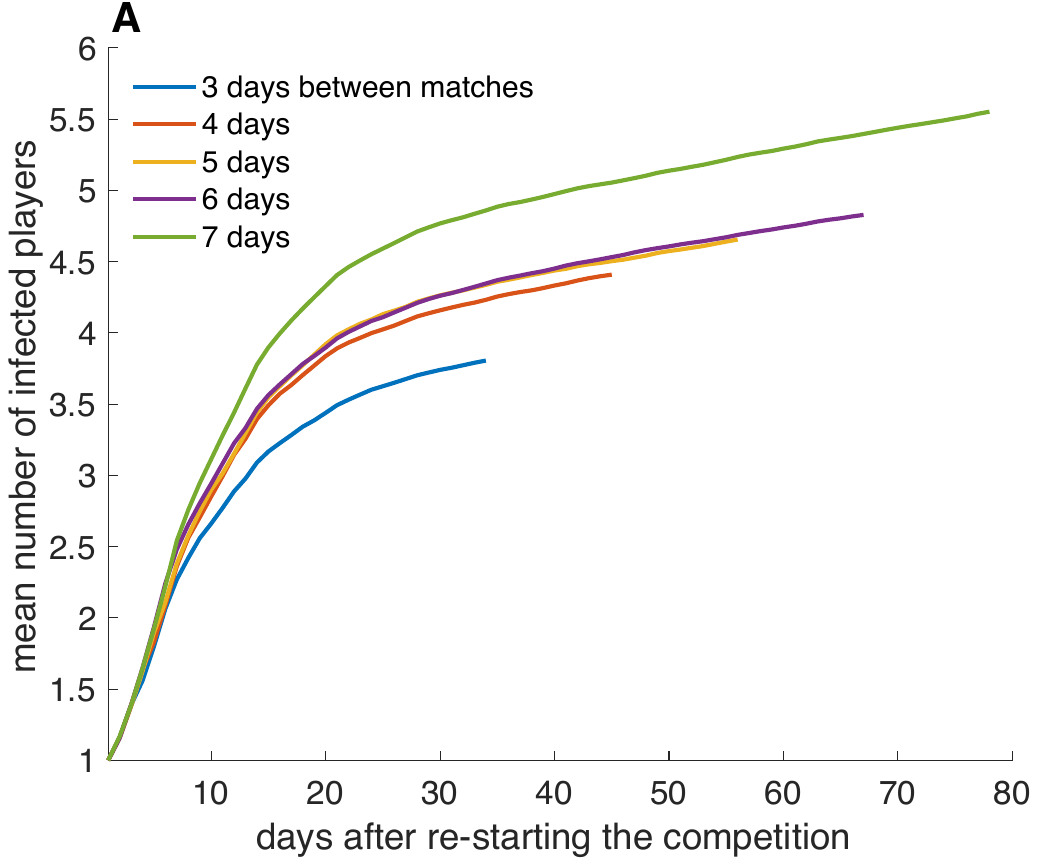}
\includegraphics[scale=0.35,angle=0,clip=]{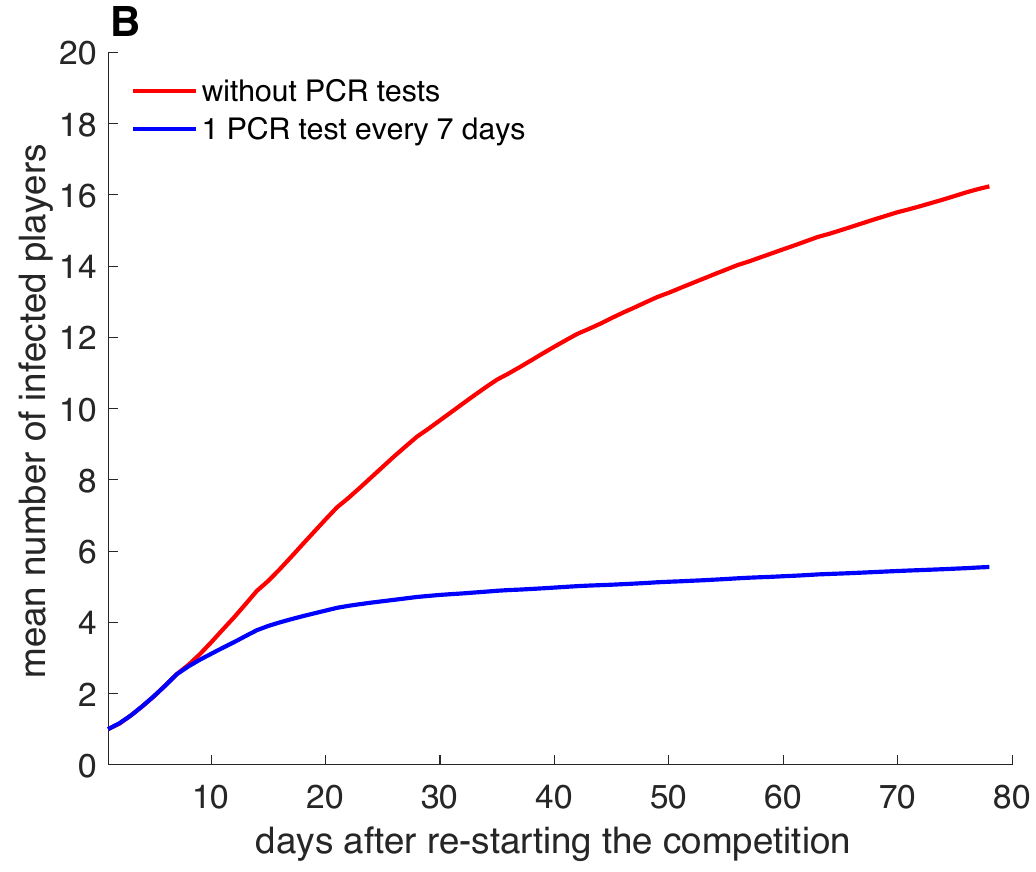}
\includegraphics[scale=0.35,angle=0,clip=]{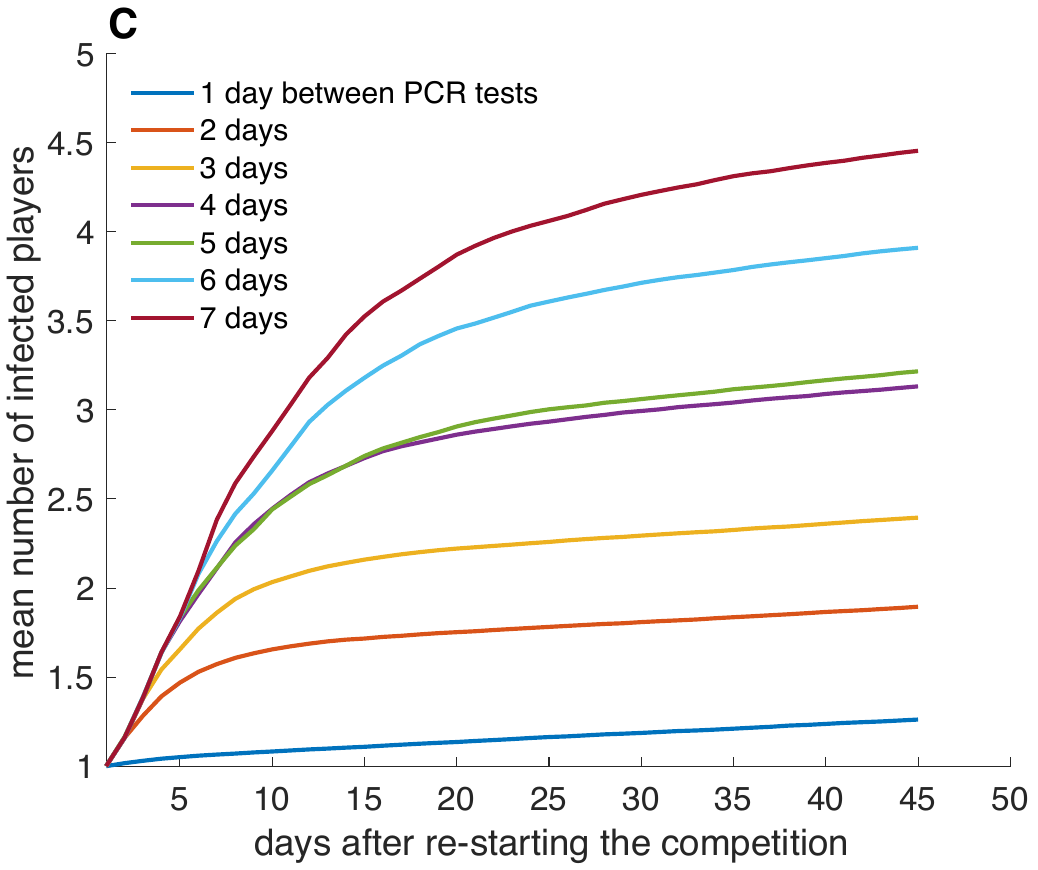}
\includegraphics[scale=0.35,angle=0,clip=]{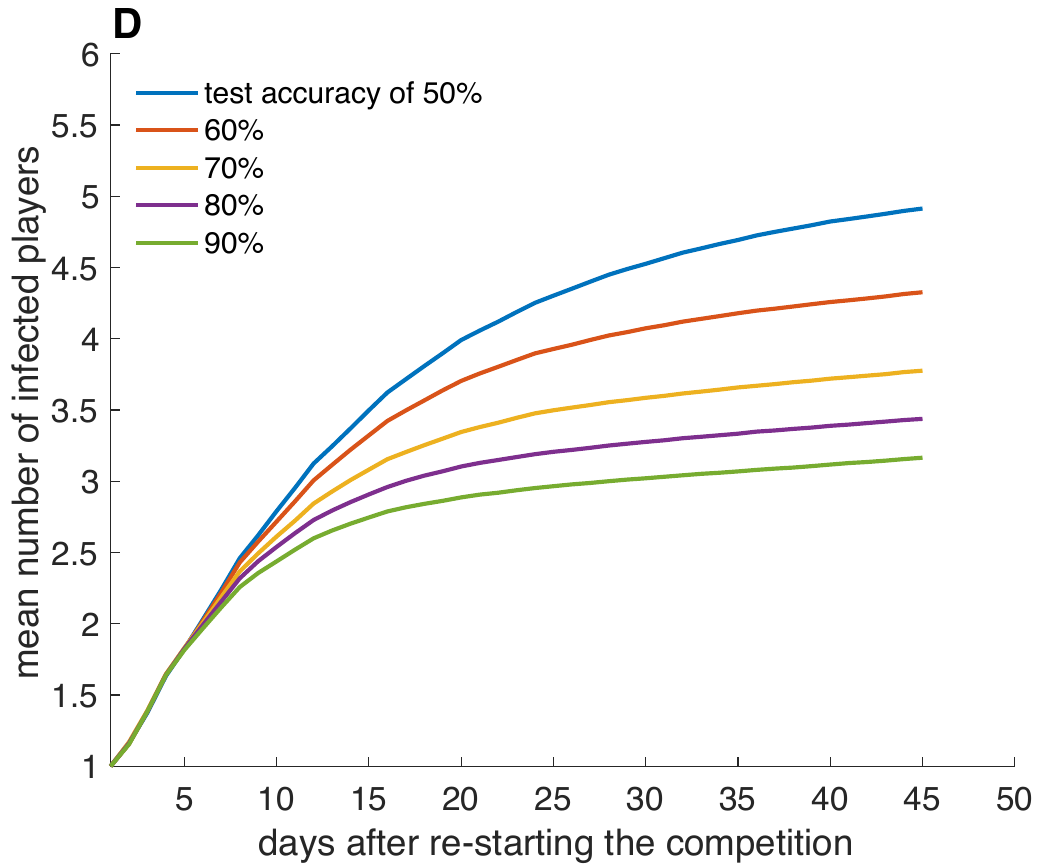}
\caption{Dependence of the mean of the accumulated number of players that have been infected at the end of the season $\bar n(t)$ 
on the main parameters of the system: 
the days between matches ($N_{match}$) (A) and the days between PCR tests ($N_{test}$) and their sensitivity ($\mu_I$) (B-D). 
We simulated $10^4$ times the rest of the season, that consists of 11 matches and the training days in between. Parameters of the simulation
are indicated in Table 1, with $\mu_I=0.9$, unless specified otherwise. The seed of all simulations contained one player 
infected at the first day of the tournament. 
(A) Influence of the number of days between two consecutive matches, $N_{match}$, on $\bar n(t)$. 
In this simulation, PCR tests with $90\%$ sensitivity were carried out every $N_{test}=7$ days.
In (B) we compare the outcome of not doing any tests during the rest of the season and doing them every $N_{test}=7$ 
days (matches played every 7 days),
while in (C) we focus on the number of days $N_{test}$ between each PCR control (matches played every 4 days, 
closer to the optimum frequency of 1 every 3 days). 
(D) Influence of test accuracy $\mu_I$ on $\bar n(t)$ (PCR tests and matches carried out every 4 days).
\label{fig02}
} 
\end{figure}

We simulated between $10^4$ and $10^5$ seasons using our discrete-time model and 
obtained the main statistics of the accumulated number of infected players at time $t$, $n(t)$.
Importantly, the seed of all simulations contained one player of the league
who is already infected at the first day of the tournament (i.e., $n(0)=1$). 
By doing so, the epidemic spreading begins at day one instead of any random day of the season,
and therefore time $t$ should be understood as days after 
the first infection.

Figure \ref{fig02} analyses the influence that the number of days between tests and matches, $N_{test}$ and $N_{match}$,
have on the accumulated number of infected players $n(t)$ along the rest of the season (i.e., 11 matches and the training days in between).
$10^4$ independent simulations were performed, and the mean values of $n(t)$, $\bar n(t)$, are plotted in the figure.

In Fig. \ref{fig02}A we see how the mean accumulated number of infected players $\bar n(t)$ changes when the number of days between 
matches $N_{match}$ is modified within the interval $\{3,4,5,6,7\}$, i.e., we set the 
minimum and the maximum number of days between matches to $3$ and $7$, respectively. 
Interestingly, we observe that it is convenient to reduce the time between matches to the minimum. 
The reason is twofold. On the one hand, with $N_{match}$ being the lowest, 
the competition would last fewer weeks, and therefore the players would be exposed for less time. 
On the other hand, the probability of being infected is higher during a training
day than during a match day, since players are more exposed to physical contact with other 
players during training. For these reasons, the higher the number of days between matches, the
higher the slope of the curves of Fig.\ref{fig02}A. 

In Fig. \ref{fig02}B we show the different evolution of the mean value of the accumulated number of infected players 
$\bar n(t)$ when PCR tests 
are or are not performed. Matches are played with a separation of $7$ days, in this case.  
We can observe how skipping the tests increases substantially the number of infected players. These results show
that conducting a coronavirus detection test is essential to prevent its spread among
{\it La Liga} teams. However, it is necessary to take into account the frequency and reliability of such tests. 
To investigate this issue, we assume that it is decided to play, for example, every 4 days, a measure close to the 
most favourable scenario of 3 days, although not so extreme. In Fig. \ref{fig02}C we see how important it 
is to perform tests as often as possible, ideally every day. As the tests are more separated over time, the risk of infecting 
more players inevitably increases. 
Finally, it is possible to simulate how important the accuracy of the tests is and the consequences of making use 
of low sensitive methods. 
Figure \ref{fig02}D shows how the mean value of infected players $\bar n(t)$ increases as the reliability of the tests $\mu_E$ and $\mu_I$ 
decreases. 
These results support
the convenience of performing PCR testing, whose accuracy is estimated to be substantially larger than any other method.

\begin{figure}[!t]
\centering
\hspace{-2.3 cm}
\includegraphics[scale=0.46,angle=0,clip=]{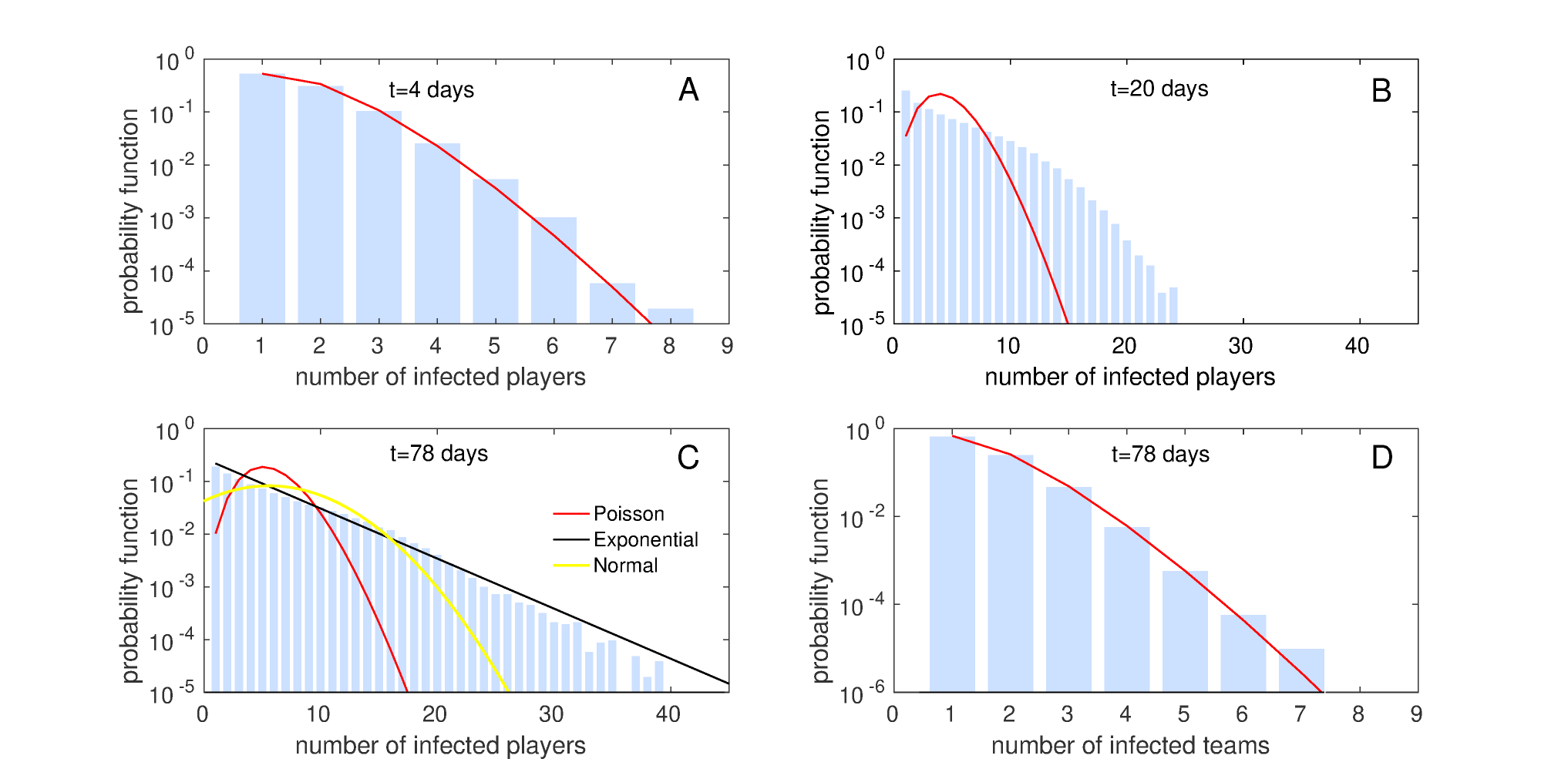}
\caption{Probability function of the accumulated number of infected
players $n(t)$ at the beginning of the re-start (A, $t=4$ days), after few weeks (B, $t=20$ days) and at the end of the season 
(C, $t=78$ days), and probability function of the accumulated number of infected
teams $M_{inf}(t)$ at the end of the season 
(D, $t=78$ days), when matches and PCR controls are carried out every 7 days. For these 4 cases, 
$\bar n \pm \sigma_n=1.6\pm 0.8 $,  $\bar n\pm \sigma_n=4.3\pm 3.6 $, $\bar n\pm \sigma_n=5.6\pm 4.8$, and
$\bar M_{inf}\pm\sigma_{M_{inf}}=1.4\pm0.6$ respectively.
Approximations to a Poisson distribution of mean
$\bar n$, to an exponential distribution of mean $\bar n$ and to a normal distribution of mean $\bar n$ and standard deviation $\sigma_n$
are shown in (A-C), while an approximation to a Poisson distribution of mean $\bar M_{inf}$ is shown in (D). $10^5$ simulations of the league were performed. 
\label{fig03}
} 
\end{figure}

As mentioned above, the curves shown in Fig. \ref{fig02} are the mean values $\bar n(t)$ obtained after $m=10^4$ simulations of the model. 
While the standard deviation of the mean $\bar n(t)$, $\sigma_{\bar{n}}=\sigma_{n}/\sqrt{m}$, is
so small that would be hardly distinguishable from the curves in any of the plots, the standard deviation of $n(t)$, $\sigma_n$, 
is on the contrary very large 
--in some cases of the order of 
the mean $\bar{n}$-- and shows that the evolution of a single process is highly unpredictable. To cast light on this question, 
in Fig. \ref{fig03} we have plotted the probability function of the accumulated number of infected
players $n(t)$ (i.e., probability of obtaining $n(t)=1,2,3...$ accumulated infected players after 
$t$ days, calculated as the normalised histogram of $10^5$ simulations of the process) 
 when matches and PCR controls are carried out every 7 days (green curve in Fig. \ref{fig02}A) after $t=4$ days, 20 days, 
and at the end of the league ($t=78$ days). In the first days of the competition (Fig. \ref{fig03}A), 
the disease starts to spread in the team of the so-far unique infected player. As expected, 
a Poisson distribution approximates accurately its function probability (note that in this and further calculations
of approximations to the data, 
we subtract the initial infected individual from 
the series and shift the obtained curve 1 position in the X-axis). However,
the disease soon spreads towards other teams and the function distribution becomes more complex: at moderate values of the time 
($t=20$, Fig. \ref{fig03}B)
the probability function presents a {\it hump} that certifies that the curve is in fact the consequence of several spreading
processes, that is, the addition of intra- and interteam spreading, plus the potential infections coming from outside the league.
Furthermore, 
when the season reaches its end ($t=78$, Fig. \ref{fig03}C) the curve presents an exponential-like tail, 
and at that time the standard deviation $\sigma_n$ is almost as large as the mean $\bar n$ (as
it is verified in exponential probability density functions). Note that, while a normal 
approximation is not accurate at the end of the league, 
when $t$ grows substantially (many weeks after the end of the season, and therefore not shown) 
the probability function becomes a Gaussian, as expected from the 
Central Limit Theorem.

Finally, let us show that the statistical behaviour of the system at the end of the season is compatible with that 
of a compound Poisson process ($CPP$), that is, 
a stochastic process with jumps, where the jumps arrive randomly according to a Poisson process.
First, the  function probability of the 
number of infected teams $M_{inf}(t)$ at $t=78$ is indeed accurately approximated by a Poisson distribution (Fig. \ref{fig03}D).
Second, the mean and standard deviation of the number of infected players at the end of the season, obtained numerically,
($\bar n=5.6$ and $\sigma_n=4.8$) agree (with an error of 6\% and 8\% respectively) 
with those obtained by the theoretical 
expressions typical of Compound Poisson processes \cite{wikipedia} for $t=78$:
\begin{eqnarray}
\bar n_{CPP}&=& 1+(\bar n_{Team}-1) \bar M_{inf}=5.9\,, \\
\sigma_{CPP}&=& \sqrt{\sigma_{n_{Team}}^2\bar M_{inf}+(\bar n_{Team}-1)^2 \sigma_{M_{inf}}^2}=5.3\,, 
\end{eqnarray}
where $n_{Team}\pm\sigma_{n_{Team}}=4.6\pm4.1$ is the accumulated number of infected players at the team where the infection started, and 
$\bar M_{inf}\pm\sigma_{M_{inf}}=1.4\pm0.6$ is the 
accumulated number of infected teams. 
In summary, in spite of the complex particular details of the model here presented and sketched in Fig. \ref{figura_modelo}, 
its statistical behaviour can 
be described as the addition of several processes, each of them happening in a different team,
and where the infection dynamics from one team to another
follows a random Poisson process.

\section{Conclusions}
\label{S:3}

{\it ``All models are wrong, but some are useful"}. This famous statement, attributed to the statistician George Box, sums up the usefulness of our model. 
Although it is not possible to predict the exact number of infected individuals, the model allows describing, in a qualitative way, the influence that different 
measures can have to mitigate the spreading of the coronavirus during a competition. Based on the simulations carried out with the 
epidemiological parameters estimated by the scientific community, the results of the study can be summarised in five points:
\begin{itemize}

\item Reducing the days between matches reduces the risk of spreading COVID-19 throughout the season. The more the season is compressed, the less risk of contagion.
 
\item PCR tests should be performed on all football players participating in the competition.  Antibody and antigen tests should be ruled out
 in this context because they are less reliable and are not effective until the disease is well advanced.

\item The tests should be carried out continuously along with the competition, with the optimum scenario being one test per day.

\item The player's environment is essential to avoid introducing the disease into the system. It is necessary that the players 
try to limit their social contacts as much as possible, and that their physical interaction with the technical staff is as distant as possible.

\item The process is highly unpredictable. While qualitative results are clear, obtaining precise predictions for a single realisation 
(the real case) is not possible. This is in agreement with recent work that warns about the strong sensitivity
to parameter values in epidemics modelling \cite{Castro:2020}.

\end{itemize}

We must also note that applying all the measures suggested by the model involves a cost. On the one hand, reducing the time 
between matches can be very physically demanding. The recovery time after a match would be reduced and 
the risk of injury would increase. To reduce this risk, teams should increase player rotations.
Regarding the tests, football clubs should provide the necessary support and means 
to carry out such a high number of tests in such a short time.
Without an adequate policy in this regard, the risk of reinfection in a competition would skyrocket.
Players will also pay a personal cost to control the eventual spreading of coronavirus.
Minimising their contacts with other individuals would mean limiting their travels, public events and, in general, 
reducing interactions with people outside the family environment. In fact, 
maintaining them in confinement during the rest of the season would be obviously the optimum situation.

Finally, although the results shown here are focused on the resumption of the men's Spanish national league,
the conclusions are equally valid for the women's competition. Furthermore, the model 
could be adapted to any competition in which matches involve
some physical contact between players, such as basketball, handball or rugby.

\section*{Acknowledgements}
The authors acknowledge E. L\'azaro and J.A. S\'anchez-Brunete for fruitful conversations on virological and pharmaceutical 
aspects of COVID-19, respectively, P. Catal\'an for advice on the statistical analysis of the results, 
and J. Iranzo
for assistance on the calculation of the model parameters. JMB acknowledges M. Casals for providing useful references about COVID-19 in sports. 
DRA is supported by Comunidad de Madrid, Spain, through project MPEJ-2019-AI/TIC-13118. JMB is funded by MINECO, Spain (FIS2017-84151-P). 
JA is supported by MINECO, Spain (FIS2017-89773-P, MiMevo), and the Spanish 
State Research Agency (AEI) Project No. MDM-2017-0737 Unidad de Excelencia 
``Mar\'ia de Maeztu''-Centro de Astrobiolog\'ia (INTA-CSIC). 









\section*{References}

\begin{thebibliography}{25}

\bibitem{who_pandemic} https://www.who.int/dg/speeches/detail/who-director-general-s-opening-remarks-at-the-media-briefing-on-covid-19---11-march-2020.

\bibitem{kermack1927} Kermack WO, McKendrick AG,
A contribution to the mathematical theory of epidemics.
Proc. R. Soc. London 1927; 115:700.

\bibitem{matadi2014} Matadi MB. 
The SIRD epidemial model. 
Far East Journal of Applied Mathematics 2014; 89;1-14.

\bibitem{li1995} Li, MY, Muldowney, JS. 
Global stability for the SEIR model in epidemiology. 
Mathematical biosciences 1995, 125; 2:155-164.

\bibitem{peng2020} Peng L, Yang W, Zhang D, Zhuge C, Hong L. 
Epidemic analysis of COVID-19 in China by dynamical modeling. 
arXiv preprint arXiv:2002.06563.

\bibitem{radulescu2020} Radulescu A,  Cavanagh K. 
Management strategies in a SEIR model of COVID 19 community spread. 
arXiv preprint arXiv:2003.11150.

\bibitem{hou2020} Hou C, Chen J, Zhou Y, Hua L, Yuan J, He S, Guo Y, Zhang S, Jia Q, Zhao C, Zhang J, Xu G, Jia E.  
The effectiveness of the quarantine of Wuhan city against the Corona Virus Disease 2019 (COVID-19): well-mixed SEIR model analysis. 
J Med Virol 2020.

\bibitem{rovetta2020} Rovetta A, Bhagavathula, AS.
Modelling the epidemiological trend and behavior of COVID-19 in Italy. 
medRxiv 2020.03.19.20038968; doi: https://doi.org/10.1101/2020.03.19.20038968

\bibitem{debrouwer2020} De Brouwer E, Raimondi D,  Moreau Y.
Modeling the COVID-19 outbreaks and the effectiveness of the containment measures adopted across countries. 
medRxiv 2020.04.02.20046375; doi: https://doi.org/10.1101/2020.04.02.20046375

\bibitem{ngonghala2020} Ngonghala CN,  Iboi E, Eikenberry S, Scotch M, MacIntyre CR, Bonds MH, Gumel AB. 
Mathematical assessment of the impact of non-pharmaceutical interventions on curtailing the 2019 novel Coronavirus. 
medRxiv 2020.04.15.20066480; doi: https://doi.org/10.1101/2020.04.15.20066480

\bibitem{lopez2020} Lopez L, Rod\'o X. 
The end of the social confinement in Spain and the COVID-19 re-emergence risk.
 medRxiv 2020.04.14.20064766; doi: https://doi.org/10.1101/2020.04.14.20064766

\bibitem{halabchi2020} Halabchi F, Ahmadinejad Z, Selk-Ghaffari M. COVID-19 Epidemic: Exercise or Not to Exercise; That is the Question! 
Asian J Sports Med. 2020;1(1):e102630.

\bibitem{duarte2020} Duarte Mu\~noz M, Meyer T. Infectious diseases and football - lessons not only from COVID-19.
Science and Medicine in Football 2020;4:85-86.

\bibitem{hay2013} Hay S. Football fever could be a dose of dengue. 
Nature 2013; 503(7477):439. 

\bibitem{corsini2020} Corsini A, Bisciotti GN, Eirale C, Volpi, P. 
Football cannot restart soon during the COVID-19 emergency! A critical perspective from the Italian experience and a call for action.
Br J Sports Med Epub ahead of print: [28/04/2020].

   \bibitem{Newman:2010} Newman M, Networks: An Introduction. Oxford University Press, Inc.; 2010.
   
  \bibitem{Erdos:1959} Erd{\"o}s P, R{\'e}nyi A. On random graphs I. Publicationes mathematicae 1959; 6(26):290--297.
  
  
   \bibitem{Wu:2020} Wu JT, Leung K, Leung GM.
   Nowcasting and forecasting the potential domestic and international spread of the 2019-nCoV outbreak originating in Wuhan, 
   China: a modelling study. The Lancet 2020; 395(10225):689--697.
   
 \bibitem{Lin:2020}  Lin Q, Zhao S, Gao D, Lou Y, Yang S, Musa S {\it et al.}  
 A conceptual model for the coronavirus disease 2019 (COVID-19) outbreak in Wuhan, China with individual reaction and governmental action.
 International Journal of Infectious Diseases 2020; 93(26):201--216.
 
  \bibitem{Lauer:2020} Lauer SA, Grantz KH, Bi Q, Jones FK, Zheng Q, Meredith HR {\it et al.} 
  The incubation period of coronavirus disease 2019 (COVID-19) from publicly reported confirmed cases: Estimation and application.
  Ann Intern Med 2020; doi: https://doi.org/10.7326/M20-0504.
  
 \bibitem{Guan:2020}  Guan W, Ni Z, Hu Y, Liang W and Ou C, He J {\it et al.} 
 Clinical Characteristics of Coronavirus Disease 2019 in China. N Engl J Med 2020; DOI: 10.1056/NEJMoa2002032.
 

   \bibitem{Wolfel:2020} W{\"o}lfel R, Corman VM, Guggemos W, Seilmaier M, Zange S, M{\"u}ller M {\it et al.} 
   Virological assessment of hospitalized patients with COVID-2019.
   Nature 2020; https://doi.org/10.1038/s41586-020-2196-x.

\bibitem{wikipedia} https://en.wikipedia.org/wiki/Compound\_Poisson\_process 

\bibitem{Castro:2020} Castro M, Ares S, Cuesta JA, Manrubia S.
Predictability: Can the turning point and end of an expanding epidemic be precisely forecast?
arXiv:2004.08842; https://arxiv.org/abs/2004.08842.



\end{thebibliography}

\end{document}